# Parallel density matrix propagation
# in spin dynamics simulations

Luke J. Edwards, Ilya Kuprov[*]

*Oxford e-Research Centre, University of Oxford,
7 Keble Road, Oxford OX1 3QG, UK.*

Email: ilya.kuprov@oerc.ox.ac.uk

Fax: +44 1865 610612

**Abstract**

Several methods for density matrix propagation in distributed computing environments, such as clusters and graphics processing units, are proposed and evaluated. It is demonstrated that the large communication overhead associated with each propagation step (two-sided multiplication of the density matrix by an exponential propagator and its conjugate) may be avoided and the simulation recast in a form that requires virtually no inter-thread communication. Good scaling is demonstrated on a 128-core (16 nodes, 8 cores each) cluster.

# 1. Introduction

In common with much of quantum theory the theoretical formalism of spin dynamics is not easily adaptable to parallel computing architectures – matrix operations in both the frequency domain (Hamiltonian diagonalization) and the time domain (density matrix propagation) have large communication overheads, resulting in poor scaling with respect to the number of CPUs on both clusters and shared-memory systems. Situations where a simulation contains multiple independent blocks (*e.g.* different orientations in a powder average or different slices in a multi-dimensional experiment) are straightforward and have already been treated in the literature[1,2] as well as implemented in mainstream simulation software packages[3,4]. However a parallel simulation of a single large spin system is considerably harder and requires alterations to the formalism as well as the simulation algorithms to expose the parallel stages and minimize the thread communication overhead.

Frequency domain simulations of large spin systems face insurmountable difficulties regardless of computer architecture – while the Hamiltonian itself is often sparse, its eigenvectors are nearly always dense meaning that a terabyte of storage (a generous allowance) would only accommodate 20 spins and any improvement would be logarithmic. Full diagonalization of Hamiltonians with dimension in excess of $10^5$ (*i.e.* with more than about 17 spins) is not at present realistic[5], although some progress has recently been made with simulations that only seek a few specific eigenvalue-eigenvector pairs[6]. Even matrix-vector multiplications can be hard to parallelize due to unpredictable fluctuations in the non-zero count between the blocks of the various array[7].

Time domain spin dynamics simulations are easier, particularly if restricted state spaces can be used[8], because small step propagators normally inherit the sparsity of the Hamiltonian[5]. The parallelization problem in the time domain is therefore reduced to finding a parallel algorithm for density matrix propagation. This issue has been investigated for optimal control problems[9], symplectic propagation of large regular spin lattices[10] and observable dynamics in spin systems[11]. The primary obstacle is that propagation steps under the Liouville - von Neumann equation involve double-sided matrix multiplication,

$$\frac{d\hat{\rho}(t)}{dt} = -i\left[\hat{H}, \hat{\rho}(t)\right] \quad \Rightarrow \quad \hat{\rho}(t+\Delta t) = e^{-i\hat{H}\Delta t}\hat{\rho}(t)e^{i\hat{H}\Delta t}, \qquad (1)$$

and so every element of $\rho(t+\Delta t)$ depends on every element of $\rho(t)$. This means that, even if elements of $\rho(t+\Delta t)$ are evaluated in parallel, the entire $\hat{\rho}(t)$ matrix has to be communicated to every node at each time step. This is an unacceptably large amount of communication – particularly on clusters, where the network bandwidth presents a bottleneck. Alternatively, the propagation step may be split using matrix factorizations (diagonalization[11] and SVD[10] have been suggested), but the factorizations are themselves expensive and difficult to parallelize.

In the present paper we propose and evaluate several methods for density matrix propagation in parallel computing environments, such as clusters and graphics processing units. It is demonstrated that the large communication overhead associated with each propagation step in Equation (1) may be avoided and the simulation recast in a form that requires no inter-thread communication beyond distributing the initial condition and retrieving the final results from the worker nodes. Good scaling is demonstrated on a 128-core (16 nodes, 8 cores each) cluster.

## 2. Computation, storage and communication overheads

The physical nature of spin interactions requires the Hamiltonian of an *n*-spin system to have at most $(n^2+3n)/2$ interactions: $(n^2-n)/2$ bilinear couplings between different spins, $n$ quadratic couplings of a spin to itself, and $n$ couplings to the external magnetic field. In the most general case, each bilinear coupling involves six linearly independent spin operators (corresponding to the isotropic part and five irreducible components of the anisotropic part), each quadratic coupling involves five operators (quadrupolar interaction and zero-field splitting are traceless, but ZFS is not necessarily axial) and each Zeeman coupling involves three spin operators ($\hat{L}_X$, $\hat{L}_Y$ and $\hat{L}_Z$). The upper bound for the number of linearly independent operators in the physically meaningful Hamiltonian of an *n*-spin system is therefore $3n^2+5n$. The fact that this number grows polynomially with the number of spins has profound algebraic and physical consequences elsewhere[8,12], but in our current context it may be used to get an upper bound on the density of the matrices (the ratio of the number of non-zeros to the total number of elements) involved in numerical spin dynamics simulations. The resulting bounds may then be used to obtain asymptotic estimates on the storage, communication and computation requirements.

It may be seen by direct inspection that matrix representations of Cartesian spin operators ($\hat{L}_X$, $\hat{L}_Y$, $\hat{L}_Z$) and their binary direct products ($2\hat{L}_X\hat{S}_X$, $2\hat{L}_X\hat{S}_Y$, *etc.*) in the Pauli basis have at most one non-zero element per row, with the number of rows for an *n*-spin system being equal to $\prod_{k=1}^{n}(2S_k+1)$, where $S_k$ is the total spin quantum number of the *k*-th spin. Of their possible combinations, the isotropic coupling operator $\hat{L}_X\hat{S}_X + \hat{L}_Y\hat{S}_Y + \hat{L}_Z\hat{S}_Z$ has at most two, and each of the five second-rank irreducible spherical tensor operators have at most one (for $\hat{T}_{2,\pm 2}$ and $\hat{T}_{2,\pm 1}$) or two (for $\hat{T}_{2,0}$) non-zeros per row. Taken together this yields an upper bound of $4n^2 + 5n$ on the number of non-zeros per row (or column) of a spin Hamiltonian and the following upper bounds on the total number of non-zeros $N_{NZ}$ and matrix density $d_{\hat{H}}$:

$$N_{NZ} \leq (4n^2 + 5n)\prod_{k=1}^{n}(2S_k + 1), \qquad d_{\hat{H}} \leq \frac{N_{NZ}}{\left[\prod_{k=1}^{n}(2S_k + 1)\right]^2} = \frac{4n^2 + 5n}{\prod_{k=1}^{n}(2S_k + 1)} \qquad (2)$$

This bound on $d_{\hat{H}}$ formally validates the common knowledge that spin operators are very sparse[13], but adds the important statement that the sparsity of the Hamiltonian matrix increases *exponentially* with the number of spins. In a 20-spin system with $S_k = 1/2$ and everything coupled to everything, $d_{\hat{H}} \leq 0.00143$ – any possible Hamiltonian would be mostly zeros.

Importantly, the sparsity analysis presented above does not apply to the density matrix. In a typical solid state NMR spin system, for instance, $\hat{\rho}$ becomes dense in the first few milliseconds of time evolution. The storage, computation and communication costs associated with the density matrix are therefore the same as for a general dense matrix. However, the cost of multiplication of $\hat{\rho}$ by the Hamiltonian scales as $O(d_{\hat{H}}N^3)$ with representation dimension $N$ – considerably better than the $O(N^3)$ cost of the dense matrix multiplication.

It is also easy to demonstrate that, for a well-chosen time step, the cost of multiplication by an exponential propagator has the same asymptotic scaling as the cost of multiplication by a Hamiltonian. Indeed, the multiplication by the exponential propagator may be expressed as

$$e^{-i\hat{H}\Delta t}\hat{\rho} = \sum_{k=0}^{\infty}\frac{(-i\hat{H}\Delta t)^k}{k!}\hat{\rho} = \sum_{k=0}^{\infty}\frac{(-i\Delta t)^k}{k!}\left(\hat{H}\left(\hat{H}\ldots(\hat{H}\hat{\rho})\right)\right). \qquad (3)$$

If the time step is chosen as $\Delta t = \|\hat{H}\|_{\infty}^{-1}$, this series converges to machine precision in about 15 iterations meaning that only a fixed number of multiplications by the Hamiltonian is in practice required. The cost of the Hamiltonian infinity-norm calculation is $O(N_{NZ})$ additions, which is negligible. The cost of forming and manipulating the Hamiltonian is similarly small compared to the cost of the same operations on the density matrix, but only if its sparsity is preserved.

From the reasoning above, we can conclude that the primary computation and communication cost of a time-domain spin dynamics simulation is associated with time propagation of the density matrix – the housekeeping costs associated with operators and propagators are negligible.

In practice this means that Hamiltonians and propagators may be replicated without loss of efficiency on distributed computing systems with up to about $1/d_{\hat{H}}$ worker nodes. The density matrix, however, clearly requires distributed storage and distributed operations; any inter-thread communication involving the density matrix must be kept to a minimum.

## 3. Expectation value dynamics for a specific observable

Parallel density matrix propagation as proposed by Skinner and Glaser[11] involves the density matrix expansion

$$\hat{\rho} = \sum_k p_k |v_k\rangle\langle v_k| \quad (4)$$

where $|v_k\rangle$ is the $k$-th eigenvector of the density matrix and $p_k$ is the corresponding eigenvalue. With this decomposition the calculation of expectation values is easy to parallelize with a relatively modest communication overhead:

$$\begin{aligned}\langle \hat{A}(t)\rangle &= \text{Tr}\left[\hat{A} e^{-i\hat{H}t} \hat{\rho}(0) e^{i\hat{H}t}\right] = \sum_k p_k \text{Tr}\left[\hat{A} e^{-i\hat{H}t} |v_k(0)\rangle\langle v_k(0)| e^{i\hat{H}t}\right] = \\ &= \sum_k p_k \left[\langle v_k(0)| e^{i\hat{H}t}\right] \hat{A} \left[e^{-i\hat{H}t} |v_k(0)\rangle\right] = \sum_k p_k \langle v_k(t)|\hat{A}|v_k(t)\rangle\end{aligned} \quad (5)$$

where the individual trajectories $e^{-i\hat{H}t}|v_k(0)\rangle$ corresponding to different values of the index $k$ are independent for the entire duration of the simulation and may be computed on different nodes. The elementary spin operators are always very sparse and the cost of supplying each node with $\hat{H}$ and $\hat{A}$ matrices is therefore acceptable. At each propagation step only the expectation values of $\hat{A}$ need to be gathered and summed on the head node, with weights specified by $\{p_k\}$. Matrix exponentiation is not required because the elementary propagation step,

$$|v_k(t+\Delta t)\rangle = e^{-i\hat{H}\Delta t}|v_k(t)\rangle, \quad (6)$$

may be computed directly from $\hat{H}$ and $|v_k(t)\rangle$ using Krylov techniques that only use sparse matrix-vector multiplications[14]. The computational scaling of Equation (5) is therefore expected to be excellent.

The expansion postulated in Equation (4) is only computationally affordable if the initial density matrix is diagonal in the current basis. If $\hat{\rho}$ is not diagonal, two kinds of problems can potentially arise: some density matrices (for example $\hat{\rho} = \hat{L}_+$) cannot be diagonalized because they are singular or near-singular, and some cannot be diagonalized because they are too large. The first problem has a simple solution – the singular value decomposition (SVD)

$$\hat{\rho} = \sum_k \sigma_k |u_k\rangle\langle v_k| \tag{7}$$

exists for any matrix and may be used instead of diagonalization in Equation (4) with the corresponding modifications applied to Equation (5):

$$\langle \hat{A}(t) \rangle = \text{Tr}\left[\hat{A}e^{-i\hat{H}t}\hat{\rho}(0)e^{i\hat{H}t}\right] = \sum_k \sigma_k \text{Tr}\left[\hat{A}e^{-i\hat{H}t}|u_k(0)\rangle\langle v_k(0)|e^{i\hat{H}t}\right] = $$
$$= \sum_k \sigma_k \left[\langle u_k(0)|e^{i\hat{H}t}\right]\hat{A}\left[e^{-i\hat{H}t}|v_k(0)\rangle\right] = \sum_k \sigma_k \langle u_k(t)|\hat{A}|v_k(t)\rangle \tag{8}$$

Because the left-side trajectory, $\langle u_k(0)|e^{i\hat{H}t}$, is no longer the conjugate of the right-side trajectory, $e^{-i\hat{H}t}|v_k(0)\rangle$, this doubles the amount of work compared to the formulation given by Skinner and Glaser[11] but avoids the problem of singular density matrices. In practical simulations the overall amount of work ends up being smaller because non-Hermitian density matrices are used to replace phase cycles, which are typically 8 to 16 simulations long. It is also often the case (particularly in weakly coupled spin systems) that the density matrix has many small singular values, which may be ignored altogether, thus reducing the amount of work in Equation (8).

While Equation (7) is in some ways an improvement on Equation (4), the matrix dimension problem still remains unsolved – for a sparse density matrix requiring $O(N)$ doubles for storage both diagonalization and SVD need $O(N^3)$ doubles to store the dense eigenvector arrays and $O(N^3)$ multiplications to obtain them[15]. Both diagonalization and SVD are also notorious for their poor parallelization, large communication overheads, and numerical accuracy issues with the degenerate eigenvalue sets that are often encountered in magnetic resonance simulations. A sparse Cholesky factorization[15] for a suitably preconditioned density matrix,

$$\hat{\rho} + \mathbb{1}\|\hat{\rho}\| = LL^\dagger \quad \Rightarrow \quad \hat{\rho}(t) = e^{-i\hat{H}t}LL^\dagger e^{i\hat{H}t} - \mathbb{1}\|\hat{\rho}\| = \left(e^{-i\hat{H}t}L\right)\left(e^{-i\hat{H}t}L\right)^\dagger - \mathbb{1}\|\hat{\rho}\|, \tag{9}$$

is similarly inefficient because it requires $O(N^3)$ multiplications on the head node, and increases matrix density (even when one reorders for sparsity[16]) thus making communications more expensive.

In view of these difficulties, we propose an alternative decomposition which does not require any kind of matrix factorization and preserves the neat parallel structure of Equation (8):

$$\hat{\rho} = \sum_k |\rho_k\rangle\langle \delta_k| \tag{10}$$

where $|\rho_k\rangle$ is the $k$-th column of the density matrix and $|\delta_k\rangle$ is a vector with 1 in position $k$ and zeros elsewhere. This formulation does not require any operations on the initial density matrix beyond sending the columns to their allocated nodes at the start of the calculation. A similar argument to Equation (8) then gives the following expression for the observable:

$$\langle \hat{A}(t) \rangle = \text{Tr}\left[ \hat{A} e^{-i\hat{H}t} \hat{\rho}(0) e^{i\hat{H}t} \right] = \sum_k \text{Tr}\left[ \hat{A} e^{-i\hat{H}t} |\rho_k(0)\rangle \langle \delta_k(0)| e^{i\hat{H}t} \right] =$$
$$= \sum_k \left[ \langle \delta_k(0)| e^{i\hat{H}t} \right] \hat{A} \left[ e^{-i\hat{H}t} |\rho_k(0)\rangle \right] = \sum_k \langle \delta_k(t)| \hat{A} |\rho_k(t)\rangle \quad (11)$$

The computational complexity of this formulation is identical to Equation (8) – two sets of vectors must be propagated at the worker nodes with no inter-thread communication apart from the negligible cost of sending the resulting observable $\langle \delta_k(t)| \hat{A} |\rho_k(t)\rangle$ back to the head node.

**Algorithm A:** expectation value dynamics for a specific observable

Step 1: distribute the columns of the initial density matrix $|\rho_k\rangle$, the columns of the unit matrix $|\delta_k\rangle$, the Hamiltonian $\hat{H}$, and the observable operator $\hat{A}$ to the worker nodes. If the step propagator $\hat{P} = \exp(-i\hat{H}\Delta t)$ is available at the start of the calculation it may be supplied to the nodes instead of the Hamiltonian.

Step 2: on each worker node $k$ pre-allocate a dense array of zeros $A^{(k)}$ for the storage of the local contribution to the observable dynamics trace.

Step 3: on each worker node $k$ propagate the local vectors and co-vectors through the prescribed number of time points and record their contribution to the observable dynamics:

$$A^{(k)}(t_n) = \langle \delta_k(t_n)| \hat{A} |\rho_k(t_n)\rangle$$
$$|\rho_k(t_{n+1})\rangle = \exp(-i\hat{H}\Delta t) |\rho_k(t_n)\rangle$$
$$|\delta_k(t_{n+1})\rangle = \exp(-i\hat{H}\Delta t) |\delta_k(t_n)\rangle$$

using matrix-vector multiplications if the exponential propagator had been supplied and Krylov propagation[14] if $\exp(-i\hat{H}\Delta t)$ is not directly available or the Hamiltonian is time-dependent.

Step 4: collect the observable traces $A^{(k)}$ from every worker node and add them up element-by-element to obtain the final observable dynamics trace.

The resulting algorithm inherits the primary advantage of the method proposed by Skinner and Glaser[11] – no inter-thread communication at the propagation stage – and also removes the need to perform any kind of matrix factorization at the problem set-up stage. The head node only needs to compute the sum of the observable traces returned by the worker nodes.

## 4. Final state calculation

The situation where the final density matrix is required after a given evolution period is similar to the calculation of observable evolution discussed in the previous section:

$$\hat{\rho}(t) = e^{-i\hat{H}t}\hat{\rho}(0)e^{i\hat{H}t} = \sum_k \left[ e^{-i\hat{H}t} |\rho_k(0)\rangle \right] \left[ \langle \delta_k(0)| e^{i\hat{H}t} \right] = \sum_k |\rho_k(t)\rangle \langle \delta_k(t)| \qquad (12)$$

The individual vectors $|\rho_k\rangle$ and $|\delta_k\rangle$ may be distributed to different worker nodes for processing and sent back to the head node at the end of the propagation (Figure 1).

**Algorithm B1:** final state calculation

  Step 1: distribute the columns of the initial density matrix $|\rho_k\rangle$, the columns of the unit matrix $|\delta_k\rangle$, and the Hamiltonian $\hat{H}$ to the worker nodes. If the step propagator $\hat{P} = \exp(-i\hat{H}\Delta t)$ is available at the start of the calculation, it may be supplied to the worker nodes instead of the Hamiltonian.

  Step 2: on each worker node $k$ propagate the local vectors and co-vectors through the prescribed number of time points:

  $$|\rho_k(t_{n+1})\rangle = \exp(-i\hat{H}\Delta t)|\rho_k(t_n)\rangle$$
  $$|\delta_k(t_{n+1})\rangle = \exp(-i\hat{H}\Delta t)|\delta_k(t_n)\rangle$$

  using matrix-vector multiplications if the exponential propagator had been supplied and Krylov propagation[14] if $\exp(-i\hat{H}\Delta t)$ is not directly available or the Hamiltonian is time-dependent.

  Step 3: collect the final vectors $|\rho_k\rangle$ and $|\delta_k\rangle$ from every worker node and re-assemble the density matrix on the head node:

  $$\hat{\rho} = \sum_k |\rho_k\rangle\langle\delta_k|$$

This algorithm is potentially less scalable because the head node has to perform more processing compared to the observable dynamics case – the cost of re-assembling the density matrix is $O(N^2)$ multiplications. It does, however, have the same advantage of zero inter-thread communication at the propagation stage.

For systems where the available communication bandwidth is large (*e.g.* shared-memory supercomputers) we would suggest a different algorithm which involves more communication but less head node processing. The basic idea is to re-order the multiplication operations in the repeated application of Equation (1) during time evolution calculations:

$$\hat{\rho}(t_n) = \underbrace{\hat{P}...\hat{P}}_{n}\hat{\rho}(0)\underbrace{\hat{P}^\dagger...\hat{P}^\dagger}_{n} = \left[\underbrace{\hat{P}...\hat{P}}_{n}\left[\underbrace{\hat{P}...\hat{P}}_{n}\hat{\rho}(0)\right]^\dagger\right]^\dagger, \qquad \hat{P} = e^{-i\hat{H}\Delta t} \qquad (13)$$

Because the density matrix is always multiplied from the left the individual columns of $\hat{\rho}(0)$ may be distributed to the worker nodes and there is no inter-thread communication during the evaluation of the inner square bracket in Equation (13). The Hermitian conjugate, however, presents a significant communication hurdle – the density matrix that was distributed column-wise between the worker nodes, and is likely to no longer be sparse, has to be transposed and re-distributed. After that an identical propagation stage is carried out and the rows of the final density matrix are sent back to the head node.

**Algorithm B2:** final state calculation

Step 1: distribute the columns of the initial density matrix $|\rho_k\rangle$ and the Hamiltonian $\hat{H}$ to the worker nodes. If the step propagator $\hat{P} = \exp(-i\hat{H}\Delta t)$ is available at the start of the calculation, it may be supplied to the worker nodes instead of the Hamiltonian.

Step 2: on each node $k$ propagate the local columns of the density matrix through the prescribed number of time points:

$$|\rho_k(t_{n+1})\rangle = \exp(-i\hat{H}\Delta t)|\rho_k(t_n)\rangle$$

using matrix-vector multiplications if the exponential propagator had been supplied and Krylov propagation[14] if $\exp(-i\hat{H}\Delta t)$ is not directly available or the Hamiltonian is time-dependent.

Step 3: re-distribute the density matrix row-wise between the worker nodes and conjugate-transpose each row on receipt to make a column.

Step 4: on each node $k$ propagate the local columns of the density matrix through the prescribed number of time points:

$$|\rho_k(t_{n+1})\rangle = \exp(-i\hat{H}\Delta t)|\rho_k(t_n)\rangle$$

using matrix-vector multiplications if the exponential propagator had been supplied and Krylov propagation[14] if $\exp(-i\hat{H}\Delta t)$ is not directly available or the Hamiltonian is time-dependent.

Step 5: collect the columns of the density matrix on the head node and conjugate-transpose the result to obtain the final density matrix.

Step 3 in this algorithm is communication-intensive and Step 5 is potentially memory-intensive on the head node – the matrix transpose operation requires a lot of random memory access, unless the matrix is stored as a sparse array[16]. However, on shared-memory architectures this may be preferred to the more memory-intensive Algorithm B1.

## 5. Performance data and summary

Table 1 gives wall clock execution times for the three parallelization methods presented in this paper on a *Matlab* cluster with 128 Intel Nehalem cores (16 nodes, 8 cores each) and 1.09 TB of RAM, provisioned from the Amazon EC2 cloud service. The numbers refers to the simulation of the detection period in a pulse-acquire NMR experiment (exact calculation with a 4096×4096 density matrix) on the 12-spin system of 3-phenylmethylene-1H,3H-naphtho-[1,8-c,d]-pyran-1-one[17] at 14.1 Tesla. In each case 35,378 time steps were taken (corresponding to 0.2 seconds of physical time), with $\Delta t \approx \left\|\hat{H}_1\right\|^{-1}$, where $\hat{H}_1$ is the rotating frame Hamiltonian. All simulations were carried out using version 1.0.959 of the *Spinach* library[5] (the source code, including the simulations discussed in this paper, is available at http://spindynamics.org).

The algorithms show satisfactory scaling all the way to 128 cores. This scaling is perfectly linear up to the node size (8 cores), but starts slowing down from 16 cores onwards due to the network communication overhead, with a factor of 2 increase in the node count yielding approximately a factor of 1.5 increase in the simulation speed. Because the node interconnect on Amazon EC2 cloud is 100BASE-TX Ethernet, these numbers should be viewed as pessimistic – typical HPC interconnects are much faster.

Of the two final state evaluation methods, the split propagation method (Algorithm B1) outperforms the double transpose method (Algorithm B2) by about 40%. This is likely a consequence of its smaller communication requirements – the split propagation method does not have the intermediate synchronization step. The calculation of observables (Algorithm A) involves an extra matrix multiplication per step compared to the final state calculation, but shows the best scaling because its communication requirements are minimal. All attempts to use density matrix factorizations (diagonalization from Equation (4), SVD from Equation (7), Cholesky factorization both with and without reordering for sparsity) as a way of splitting the propagation problem have led to much slower algorithms. It is clear that all forms of density matrix factorizations must be avoided and thus Equation (10) is the best way forward.

Magnetic resonance simulations have another frequently occurring calculation type where the full density matrix is stored at each step. Despite putting considerable effort into the matter, we were unable to find a way of parallelizing this trajectory calculation in a way that would exhibit acceptable scaling on either clusters or shared-memory systems – the number of density matrices to be communicated over the interconnect equals the number of time steps, meaning that communication becomes a major bottleneck. Another problem is memory: calculation of the full system trajectory for the example simulation described above would require a terabyte of storage. Significant compression may be achieved by using difference encoding (the difference from the previous density matrix may be stored, rather than the full density matrix, at each step), but the communication overhead of a trajectory calculation is still unacceptably large and further work is certainly required in this direction.

In summary, we found that it is possible to cast the density matrix propagation problem in a way that avoids inter-thread communication and thus enables large-scale parallel processing. To achieve satisfactory scaling on systems with 8 or more processor cores density matrix factorizations (diagonalization, singular value decomposition and Cholesky factorization were attempted in this work) must be avoided. This is possible (Equations (10) and (11) demonstrate the procedure) and enables efficient parallel calculation of final states and observable dynamics on supercomputers with up to 128 processor cores in our test calculations.

**Acknowledgements**

The authors are grateful to Tom Skinner, Steffen Glaser and Stef Salvini for helpful discussions. The project is funded by the EPSRC (EP/F065205/1, EP/H003789/1) and supported by the Oxford e-Research Centre.

**References**

(1) V. Y. Orekhov, V. A. Jaravine, *Progr. NMR Spec.* **2011**, *59*, 271-292.

(2) J. H. Kristensen, I. Farnan, *J. Magn. Reson.* **2003**, *161*, 183-190.

(3) Z. Tosner, R. Andersen, N. C. Nielsen, T. Vosegaard, *SIMPSON (version 3.1)*, 2011.

(4) M. Veshtort, R. G. Griffin, *J. Magn. Reson.* **2006**, *178*, 248-282.

(5) I. Kuprov, H. J. Hogben, M. Krzystyniak, G. T. P. Charnock, P. J. Hore, *J. Magn. Reson.* **2011**, *208*, 179-194.

(6) A. Weiße, H. Fehske, in *Computational many-particle physics*; Springer: 2008; Vol. 739, p 529-544.

(7) A. Buluc, J. R. Gilbert, in *Proceedings of the 37$^{th}$ International Conference on Parallel Processing*; IEEE: 2008, p 503-510.

(8) A. Karabanov, I. Kuprov, G. T. P. Charnock, A. v. d. Drift, L. J. Edwards, W. Köckenberger, *J. Chem. Phys.* **2011**, *135*, 084106.

(9) T. Gradl, A. Sporl, T. Huckle, S. J. Glaser, T. Schulte-Herbruggen, *Lect. Notes Comp. Sc.* **2006**, *4128*, 751-762.

(10) C. H. Woo, P. W. Ma, *Phys. Rev. E* **2009**, *79*.

(11) T. Skinner, S. Glaser, *Phys. Rev. A* **2002**, *66*.

(12) I. Kuprov, N. Wagner-Rundell, P. J. Hore, *J. Magn. Reson.* **2007**, *189*, 241-250.

(13) R. S. Dumont, S. Jain, A. Bain, *J. Chem. Phys.* **1997**, *106*, 5928-5936.

(14) M. Hochbruck, C. Lubich, *SIAM J. Numer. Anal.* **1997**, *34*, 1911-1925.

(15) G. H. Golub, C. F. Van Loan, *Matrix computations*; 3rd ed.; Johns Hopkins University Press, 1996.

(16) J. R. Gilbert, C. Moler, R. Schreiber, *SIAM J. Mat. Anal. Appl.* **1992**, *13*, 333-356.

(17) P. N. Penchev, N. M. Stoyanov, M. N. Marinov, *Spectrochim. Acta A* **2011**, *78*, 559-565.

**Figure captions**

**Figure 1.**  Schematic illustration of the split propagation method (Algorithm B1) for the evaluation of final state. The density matrix is factorized according to Equations (4), (7) or (10), the two factors are sliced and propagated independently. The matrix is reassembled at the head node at the end of the calculation. The best performance is in practice achieved with Equation (10) that avoids computationally expensive factorizations.

**Figure 2.**  Schematic illustration of the double transpose method (Algorithm B2) for the evaluation of final state. The density matrix is sliced and propagated under the left side propagator, then transposed, redistributed and propagated under the right side propagator. Compared to the split propagation method, this algorithm requires less processing the head node, but has greater communication requirements.

**Table 1.** Scaling behaviour of the parallel propagation algorithms.

| Number of CPU cores | Time steps per wall clock second | | |
|---|---|---|---|
| | Algorithm A (observable) | Algorithm B1 (final state) | Algorithm B2 (final state) |
| 1 | 1.2 | 3.1 | 1.9 |
| 2 | 2.5 | 6.2 | 3.7 |
| 4 | 4.9 | 12.5 | 7.4 |
| 8 | 9.9 | 25.1 | 14.8 |
| 16 | 18.9 | 49.7 | 29.8 |
| 32 | 29.4 | 72.7 | 48.1 |
| 64 | 48.4 | 112.8 | 78.6 |
| 128 | 68.0 | 151.7 | 110.9 |

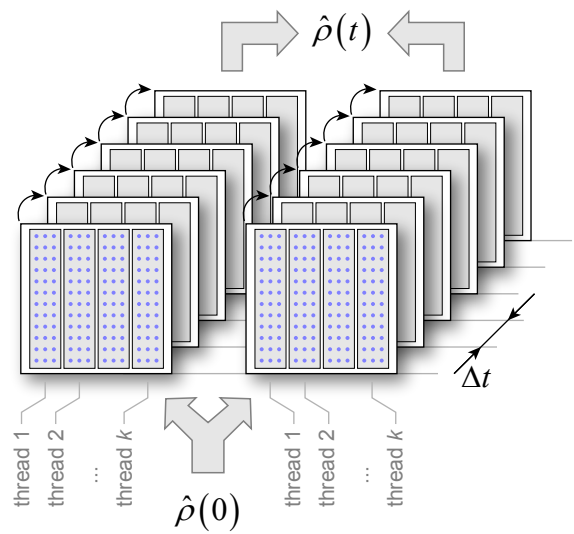

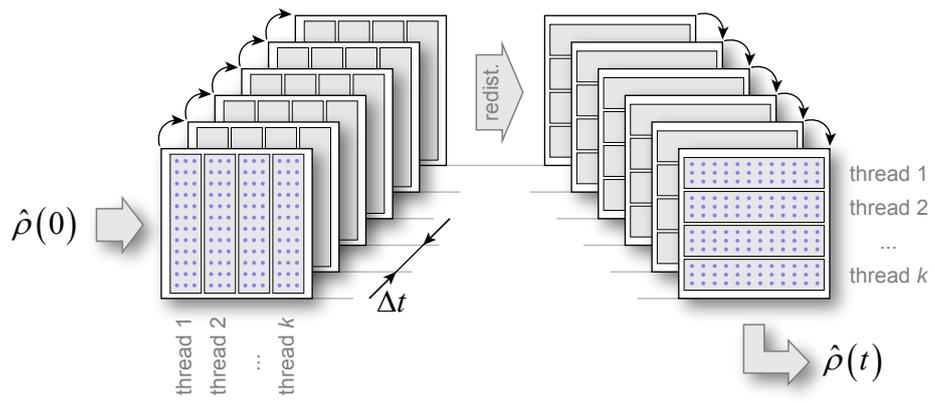